\documentclass[prd,aps,12pt,nofootinbib,superscriptaddress,showpacs]{revtex4-2}

\usepackage{amsmath}

\usepackage{hyperref}
\hypersetup{
	colorlinks=true,
	allcolors=blue,
	linktocpage=true,
	pdfhighlight=/N
}
\usepackage{bookmark}

\begin{document}

\title{Light scalaron as dark matter}

\author{Yuri Shtanov}
\affiliation{Bogolyubov Institute for Theoretical Physics,  Metrologichna St.~14-b, Kiev 03143, Ukraine} %
\affiliation{Taras Shevchenko National University of Kiev, Volodymyrska St.~60, Kiev 01033, Ukraine} %

\begin{abstract}
A new cosmological scenario is proposed in which a light scalaron of $f (R)$ gravity plays the role of dark matter. In this scenario, the scalaron initially resides at the minimum of its effective potential while the electroweak symmetry is unbroken. At the beginning of the electroweak crossover, the evolving expectation value of the Higgs field triggers the evolution of the scalaron due to interaction between these fields. After the electroweak crossover, the oscillating scalaron can represent cold dark matter. Its current energy density depends on a single free parameter, the scalaron mass $m$, and the value $m \simeq 4 \times 10^{-3}\, \text{eV}$ is required to explain the observed dark-matter abundance.  Larger mass values would be required in scenarios where the scalaron is excited before the electroweak crossover.
\end{abstract}

\maketitle

\section{Introduction}

Several researchers explored the possibility that dark matter in our universe can be explained in frames of $f (R)$ gravity models \cite{Nojiri:2008nt, Cembranos:2008gj, Corda:2011aa, Katsuragawa:2016yir, Katsuragawa:2017wge, Yadav:2018llv, Parbin:2020bpp} (more suggestions in this direction can be found in reviews \cite{Sotiriou:2008rp, DeFelice:2010aj}). In these models, dark matter is associated with the scalaron field that arises when proceeding from the Jordan frame to the Einstein frame by a conformal transformation of the metric. An appealing feature of this approach is that a dark-matter candidate arises here purely from the gravitational sector, without introducing new fundamental fields (although introducing a new degree of freedom in this sector).

In most papers on the subject, the ordinary matter is treated macroscopically and is characterised by averaged energy density and pressure in the Jordan frame. The scalaron potential then depends on the matter density, varying in time and over different astrophysical objects (so-called ``chameleon'' effect). In this paper, we assume the usual description of matter by the fundamental Standard-Model action in the Jordan frame, and proceed to the Einstein frame taking into account the conformal properties of this action. One thus obtains a model with the Einstein gravity and with a non-trivial interaction between the scalaron and the Higgs field, which can be taken as the starting point. 

Based on such a model, we propose a new cosmological scenario with the scalaron playing the role of dark matter.  In this scenario, the scalaron is assumed to reside initially at the minimum of its effective potential while the electroweak symmetry is unbroken. At the beginning of the electroweak crossover, the evolving expectation value of the Higgs field makes the scalaron also to evolve. After the electroweak crossover, the oscillating scalaron can play the role of cold dark matter.  Its current energy density in this scenario depends on a single free parameter, the scalaron mass $m$, and the value $m \simeq 4 \times 10^{-3}\, \text{eV}$ is then required to explain the observed dark-matter abundance.  

\section{The model}

For definiteness, we consider the action for gravity of the form 
\begin{equation} \label{Sg}
S_g = - \frac{M^2}{3} \int d^4 x \sqrt{-g} \left (2 \Lambda + R - \frac{R^2}{6 m^2} \right) \, ,
\end{equation} 
where $M = \sqrt{3 / 16 \pi G} \approx 3 \times 10^{18}\, \text{GeV}$ is a conveniently normalised Planck mass, and $\Lambda \approx \left( 3 \times 10^{-33}\,\text{eV} \right)^2$ is the cosmological constant in the natural units $\hbar = c = 1$. 

In the original model due to Starobinsky \cite{Starobinsky:1980te}, one usually sets $m \simeq 10^{-5} M$ in (\ref{Sg}) to account for the inflationary stage leading to predictions for the primordial power spectrum remarkably consistent with current observations \cite{Akrami:2018odb}. However, similarly to the proposal made in \cite{Cembranos:2008gj}, we would like to use this model to describe dark matter, and we are going to allow the $R^2$ correction to enter with a large constant in front of it, or, in other words, to allow $m$ to be much smaller than the cited value.  The smallness of $m$ might seem unnatural but, if we consider pure gravity and express the Lagrangian in (\ref{Sg}) in terms of the dimensionless scalar curvature ${\cal R} = R / 2 \Lambda$, then it takes the form
\begin{equation}\label{Lg}
L_g = - \frac{2 M^2 \Lambda}{3} \left( 1 + {\cal R} - \frac{\Lambda}{3 m^2} {\cal R}^2 \right) \, .
\end{equation}
For not too small values of $m$, namely, for $m \gg 10^{-33}\, \text{eV}$, the term  quadratic in the dimensionless curvature appears here with a small dimensionless factor.  Of course, this is a manifestation of the extreme smallness of the cosmological constant, formally the leading term in the expansion of (\ref{Sg}) in powers of curvature. The smallness of both $\Lambda$ and $m$ can be regarded as a naturalness (hierarchy) issue of particle physics, which is not resolved in this paper. 

We proceed from the Jordan frame of fields to the Einstein frame by the usual conformal transformation
\begin{equation} \label{om}
g_{\mu\nu} \to \Omega^{-2} g_{\mu\nu}
\end{equation}
with $\Omega = e^{\phi / 2 M}$, where $\phi$ is a new field (the scalaron). Action (\ref{Sg}) then becomes the Einstein action with the scalar field $\phi$:
\begin{align}\label{Sgn}
S_g &= - \frac{M^2}{3} \int d^4 x \sqrt{-g} R \nonumber \\ &\quad + \int d^4 x \sqrt{-g} \left[ \frac12 g^{\mu\nu} \partial_\mu \phi \partial_\nu \phi - \frac12  m^2 M^2 \left( 1 - e^{- \phi / M} \right)^2 - \frac23 \Lambda M^2 e^{- 2 \phi / M} \right] \, . 
\end{align}

Consider now the matter part of the action. We assume that it has the usual form of the Standard Model in the Jordan frame. Proceeding to the Einstein frame affects this action as well. Note, however, that most of the Standard Model action is classically conformally invariant (with proper conformal transformation of the matter fields), and, therefore, will retain its original form. The only part that breaks conformal invariance is the Higgs sector, with the action
\begin{equation}\label{Sh}
S_h = \int d^4 x \sqrt{-g} \left[ g^{\mu\nu} \left( D_\mu \Phi \right)^\dagger D_\nu \Phi - \lambda \left( \Phi^\dagger \Phi - \frac{v^2}{2} \right)^2 \right] \, .
\end{equation}
Here, $D_\mu$ is the gauge covariant derivative involving the SU(2) and U(1) electroweak gauge fields and acting on the Higgs doublet $\Phi$, and
$v \approx 246\, \text{GeV}$ is the symmetry-breaking parameter. After the conformal transformation (\ref{om}) accompanied by the Higgs-field transformation $\Phi \to \Omega \Phi$, this action becomes\footnote{Similar result of the conformal transformation of fields was under discussion, e.g., in \cite{Burrage:2018dvt}.}
\begin{align}\label{Shn}
S_h = \int d^4 x \sqrt{-g} g^{\mu\nu} \left[ \left( D_\mu \Phi \right)^\dagger D_\nu \Phi + \frac{1}{2 M} \partial_\mu \left( \Phi^\dagger \Phi \right) \partial_\nu \phi + \frac{1}{4 M^2} \Phi^\dagger \Phi\, \partial_\mu \phi \partial_\nu \phi \right] \nonumber \\ {} - \int d^4 x \sqrt{-g} \lambda \left( \Phi^\dagger \Phi - e^{- \phi / M} \frac{v^2}{2} \right)^2 \, . 
\end{align}
We observe the appearance of non-renormalisable interactions of the scalaron $\phi$ with the Higgs field in (\ref{Shn}), which, however, are all suppressed by inverse powers of the large Planck mass $M$.

The scalaron and the Higgs field are slightly mixed in this model. Choosing the canonical unitary gauge for the Higgs doublet $\Phi$ and shifting it by its vacuum expectation value $v$, we write
\begin{equation}
\Phi = \frac{1}{\sqrt{2}} \begin{pmatrix} 0 \\ h \end{pmatrix} = \frac{1}{\sqrt{2}} \begin{pmatrix} 0 \\ v + \varphi \end{pmatrix} \, .
\end{equation}
where $\varphi$ is the shifted real-valued Higgs field. The quadratic part of the Lagrangian in (\ref{Sgn}) and (\ref{Shn}) is then (in what follows, we neglect the small cosmological constant)
\begin{equation}\label{L2}
L_2 = \frac12 \left( \partial \varphi \right)^2 + \frac12 \left( 1 + \frac{v^2}{4 M^2} \right) \left( \partial \phi \right)^2 + \frac{v}{2 M} \partial \varphi \partial \phi - \lambda v^2 \varphi^2 - \frac12 m^2 \phi^2 - \frac{\lambda v^4}{4 M^2} \phi^2 - \frac{\lambda v^3}{M} \varphi \phi \, .
\end{equation}
Introducing a modified Higgs field $\chi = \varphi + {v \phi}/{2 M}$, one brings (\ref{L2}) to the diagonal form
\begin{equation}
L_2 = \frac12 \left( \partial \chi \right)^2 + \frac12 \left( \partial \phi \right)^2 - \lambda v^2 \chi^2 - \frac12 m^2 \phi^2 \, .
\end{equation}
In the Standard Model, the Higgs field has mass $m_\text{H} = \sqrt{2 \lambda} v \approx 125\,\text{GeV}$, so that $\lambda \approx 0.13$.

We can take the Einstein frame of fields as our starting frame in which all fields are to be quantised. In this frame, initially we are dealing only with the scalaron--Higgs-field interaction described by (\ref{Shn}). However, one can also start with the original Jordan frame; in this case, in the Einstein frame there arise additional interactions between the scalaron and gauge fields caused by the quantum anomalous character of conformal transformation of fields \cite{Cembranos:2008gj, Katsuragawa:2016yir}. Such interactions are also suppressed by inverse powers of the large Planck mass $M$. Their presence will not affect our scenario in any essential way.  Some of their potential effects are discussed in the last section. 

\section{The scenario}

In thermal equilibrium, the Higgs field acquires thermal corrections to its potential, which, for large temperatures $T \gtrsim m_i$, where $m_i$ are particle masses caused by interaction with the Higgs field, can be estimated as (see \cite{GR})
\begin{equation}\label{VT}
V_T ( h ) \simeq \frac16 T^2 h^2 - \frac{1}{100} T h^3 \, .
\end{equation}
Collecting also the relevant parts in (\ref{Sgn}) and (\ref{Shn}), we obtain the effective Lagrangian for the Higgs $h$ and the scalaron $\phi$:
\begin{align}\label{Leff}
L_\text{eff} =  \frac12 \left( \partial \phi \right)^2 + \frac12 \left( \partial h \right)^2 + \frac{1}{2 M} h\, \left( \partial h \partial \phi \right) + \frac{1}{8 M^2} h^2 \left( \partial \phi \right)^2 \nonumber \\ {} - \frac12 m^2 M^2 \left( 1 - e^{- \phi / M} \right)^2 - \frac{\lambda}{4} \left( h^2 - e^{- \phi / M} v^2 \right)^2 - \frac16 T^2 h^2 + \frac{1}{100} T h^3 \, .
\end{align}
Here we neglect the vacuum quantum corrections to the Higgs effective potential.

As regards the Higgs field, which strongly interacts with the primordial plasma, it is assumed to be in thermal equilibrium, hence, its expectation value $h$ residing in the minimum of its effective potential in (\ref{Leff}). The cosmological equation for the spatially homogeneous scalaron is
\begin{align} \label{eqs}
\ddot \phi + 3 H \dot \phi + \frac{1}{4 M} \left[ \left( h^2 \right)^{\cdot\cdot} + 3 H \left( h^2 \right)^{\cdot} \right] + \frac{1}{4 M^2} \left[ \left( h^2 \dot \phi \right)^\cdot + 3 H h^2 \dot \phi \right] \nonumber \\ {} + m^2 M e^{- \phi / M} \left( 1 - e^{- \phi / M} \right) + \frac{\lambda v^2}{2 M} e^{- \phi / M} \left( h^2 - e^{- \phi / M} v^2 \right) = 0 \, . 
\end{align}
Here, $H = \dot a /a$ is the Hubble parameter, with $a$ being the scale factor in the space-time metric of the expanding universe. 

In this work, we consider a scenario in which the scalaron initially resides at the minimum of its potential in (\ref{Leff}) before the electroweak crossover. The reason for this could be that its interactions with the Higgs field, present in (\ref{Shn}) or (\ref{Leff}), being strong at very large energies (very high temperatures), have relaxed the scalaron to the minimum of its potential.  In any case, this is definitely a special initial condition for the scalaron.

The value of $\phi$ at the minimum of its potential in (\ref{Leff}) is given by
\begin{equation}
\phi_0 (h) = M \ln \left( \frac{2 m^2 M^2 + \lambda v^4}{2 m^2 M^2 + \lambda v^2 h^2} \right) \, .
\end{equation}
Thus, at $h = 0$ (before the electroweak crossover), we obtain the initial value
\begin{equation}\label{phini}
\phi_0 = M \ln \left( 1 + \frac{\lambda v^4}{2 m^2 M^2} \right) \, . 
\end{equation}

The temperature $T_c$ of the beginning of the electroweak crossover is determined by vanishing of the second derivative of the effective potential in (\ref{Leff}) with respect to $h$ at $h = 0$. Performing a simple calculation, one obtains
\begin{equation}\label{Tc}
T_c \simeq \sqrt{3 \lambda} v e^{- \phi_0 / 2 M}  = \frac{\sqrt{3 \lambda} v}{\sqrt{1 + \lambda v^4 / 2 m^2 M^2}} \, .
\end{equation}

We assume the inequality\footnote{The opposite inequality does not lead to reasonable cosmology, as discussed in Appendix~\ref{sec:smass}.}
\begin{equation} \label{ineq2}
\frac{\lambda v^4}{2 m^2 M^2} \ll 1 \, ,
\end{equation}
which implies $m \gg 5 \times 10^{-6}\, \text{eV}$.  In this case, the electroweak crossover proceeds in the usual way, starting at temperature [see (\ref{Tc})]
\begin{equation}
T_c \simeq \sqrt{3 \lambda} v \approx 154\, \text{GeV} \, .
\end{equation}
At this moment, according to (\ref{phini}) and (\ref{ineq2}), 
\begin{equation}\label{phi0}
\phi_0 \approx \frac{\lambda v^4}{2 m^2 M} \ll M  \, .
\end{equation}

After the beginning of electroweak crossover, the Higgs field relaxes to the minimum of its effective potential [we neglect the last term in (\ref{Leff}), which is legitimate at early stages of crossover]:
\begin{equation} \label{relax}
h^2 = v^2 e^{- \phi  / M} - \frac{T^2}{3 \lambda} \approx v^2 \left( 1 - \frac{T^2}{T_c^2} \right) \, ,
\end{equation}
where we have used the condition $|\phi| / M \ll 1$.  From this moment on, the scalaron starts evolving according to equations (\ref{eqs}) with $h^2$ given by (\ref{relax}). Such an approximation to the field dynamics can be justified by the fact that the relaxation time of the Higgs field is much smaller than the scalaron time scale $m^{-1}$. Indeed, the Higgs effective mass $m_\text{H} (h) = \sqrt{\lambda} h$ during the crossover becomes much larger than $m$ for $h / v \gg m / \sqrt{\lambda} v \sim  m / 10^{2}\, \text{GeV}$, i.e., practically at the very beginning of crossover for small scalaron masses $m \ll 10^{2}\, \text{GeV}$.

Several terms in equation (\ref{eqs}) can be neglected. Thus, when compared to the scalaron mass term $m^2 \phi$, all terms with time derivatives in (\ref{eqs}) containing the Higgs field can be dropped because of the presence of a very small parameter ${v^2}/{M^2} \sim 10^{-32}$. Therefore, we can use the following simple equation for the scalaron:
\begin{equation} \label{eqsf}
\ddot \phi + 3 H \dot \phi + m^2 \phi + \frac{\lambda v^2}{2 M} \left( h^2 - v^2 \right) = 0 \, , 
\end{equation}
in deriving which, we have neglected terms of higher order in $|\phi| / M \ll 1$.

The value of the Hubble parameter at the electroweak crossover is given by
\begin{equation}\label{Hc}
H_c \equiv \left( \frac{\rho_c}{2 M^2} \right)^{1/2} \equiv \left( \frac{\pi^2 g_c T_c^4}{60 M^2} \right)^{1/2} \simeq \left( \frac{3 \pi^2 g_c}{20} \right)^{1/2} \frac{\lambda v^2}{M} \simeq 3 \times 10^{-5}\, \text{eV} \, .
\end{equation}
Here, $g_c \approx 100$ is the number of relativistic degrees of freedom  in thermal equilibrium at this moment. We anticipate (this will follow from our final estimates) that $H_c \ll m$, hence, that the Hubble friction term in (\ref{eqsf}) is small compared to the scalaron mass term starting from the electroweak crossover. This even strengthens inequality (\ref{ineq2}).

Introducing the dimensionless ratio $\xi = {\phi} / {\phi_0}$, one can write equation (\ref{eqsf}) in the form
\begin{equation}\label{eqxi}
\ddot \xi + 3 H \dot \xi + m^2 \left( \xi - \xi_h \right) = 0 \, ,
\end{equation}
where $\xi_h = 1 - h^2 / v^2$. The quantity $\xi_h$ decreases from unity at the beginning of electroweak crossover to the zero value asymptotically. At earlier stages of the electroweak crossover, according to (\ref{relax}), it evolves as $\xi_h = T^2/T_c^2$. Later on, it deviates from this law because of deviation of the Higgs thermal potential from the high-temperature approximation (\ref{VT}). In particular, the leading contributions to the effective potential for the Higgs field during crossover contain exponential factors of the form $e^{- m_i / T}$, where $m_i$ are the (temperature-dependent) particle masses \cite{Laine}. The expectation value of the Higgs field at temperatures $T \sim m_i$ will decrease with the rate of change 
\begin{equation}
t_h^{-1} \sim \frac{|\dot T| m_i}{T^2} \simeq \frac{H m_i}{T} \, .
\end{equation}
This inverse time scale is of the order $H_c$ at the beginning of crossover, decreasing as $t_h^{-1} \propto T$ with temperature. Thus, if $m t_h \sim m / H_c \gg 1$ initially (which we assumed to be the case), then it remains to be large all the time. Since we also have $H \ll m$, the evolution of the quantity $\xi$ is adiabatic with respect to the mass $m$ after the beginning of crossover: according to (\ref{eqxi}), it will oscillate around the slowly changing value $\xi_h$ with slowly decreasing amplitude.

Such an evolution is characterised by the adiabatic invariant of the damped oscillator described by the deviation $\Delta = \xi - \xi_h$:
\begin{equation} \label{adiabat}
a^3 E = \frac{a^3}{2} \left( \dot \Delta^2 + m^2 \Delta^2 \right) = \text{const} \, .
\end{equation} 
As the electroweak crossover terminates, the quantity $\xi_h$ relaxes to zero, and the energy $E$ then represents the energy of the variable $\xi$ itself.  These results are substantiated in Appendix~\ref{sec:solution}, where an approximate solution of equation (\ref{eqxi}) is presented for the case $\xi_h = T^2 / T_c^2$.
 
The initial value $E_c$ is determined by the initial condition $\xi = 1$, $\dot \xi = 0$ at $T = T_c$. This condition arises as a result of non-adiabatic commencement of the electroweak crossover, in which the time derivative of $\xi_h$ develops from zero to its initial value $\dot \xi_h = - 2 H_c \xi_h = - 2 H_c$ in a time much smaller than $m^{-1}$ (i.e., instantaneously, compared to the time scale of evolution of the scalaron). This is justified by the condition that the relaxation time of the Higgs field is much smaller than $m^{-1}$.

Thus, at the beginning of crossover (at $T = T_c$), we have $\Delta = 0$ and $\dot \Delta = - \dot \xi_h = 2 H_c$, so that $E_c = 2 H_c^2$, and the energy of oscillations evolves according to (\ref{adiabat}) as
\begin{equation} \label{energy}
E = 2 H_c^2 \left( \frac{a_c}{a} \right)^3 \, .
\end{equation}
Multiplication by $\phi_0^2$ then gives us the late-time behaviour of the scalaron energy density:
\begin{equation}
\rho_\phi = 2 \phi_0^2 H_c^2 \left( \frac{a_c}{a} \right)^3 = \frac{2 \pi^2 \phi_0^2 g_c T_c^4}{15 M^2} \left( \frac{a_c}{a} \right)^3 \, .
\end{equation}

For the present epoch with the scale factor $a_0$, we have $ \left( {a_c}/{a_0} \right)^3 = {2 T_0^3}/{g_c T_c^3}$, where $T_0 = 2.34 \times 10^{-4}\, \text{eV}$ is the temperature of the cosmic microwave background. Therefore, the present energy density of the scalaron is
\begin{equation}
\rho_0 = \frac{4 \pi^2 \phi_0^2 T_c T_0^3}{15 M^2} \simeq \frac{\pi^2 \lambda^{5/2} v^9 T_0^3}{5 \sqrt{3} m^4 M^4} \,  ,
\end{equation}
and the corresponding cosmological parameter is
\begin{equation}
\Omega_\phi = \frac{\rho_0}{2 M^2 H_0^2} \simeq \frac{\pi^2 \lambda^{5/2} v^9 T_0^3}{10 \sqrt{3} m^4 M^6 H_0^2} \,.
\end{equation}
Substituting here the values of the physical quantities, we obtain
\begin{equation} \label{Om}
\Omega_\phi h_{100}^2 \simeq 0.12 \left( \frac{4.4 \times 10^{-3} \, \text{eV}}{m} \right)^4 \, ,
\end{equation}
where $h_{100} = H_0 / 100\, \text{km}\,\text{s}^{-1}\text{Mpc}^{-1}$. Thus, for $m$ of the indicated magnitude, we obtain the result consistent with the established \cite{Aghanim:2018eyx} abundance of dark matter, $\Omega_c h_{100}^2 = 0.12$.  

\section{Discussion}

In this letter, we proposed a new scenario in which a light scalaron in the $f(R)$ gravity in the form (\ref{Sg}) plays the role of dark matter. Starting from the Standard Model action in the Jordan frame, we proceeded to the Einstein frame obtaining theory (\ref{Shn}), in which the scalaron interacts non-trivially with the Higgs field. The scalaron is then assumed to reside initially at the minimum of its effective potential while the electroweak symmetry is unbroken.  The subsequent evolution of the Higgs field during the electroweak crossover triggers the evolution of the scalaron.\footnote{Triggering the scalaron oscillations by the abrupt radiative breaking of the scale and electroweak symmetries was discussed in \cite{Katsuragawa:2018wbe} in the context of the ``chameleon'' mechanism in the Jordan frame.} It starts oscillating with decreasing amplitude around its slowly changing stationary point and, eventually, plays the role of dark matter if its mass is about $m \simeq 4 \times 10^{-3}\, \text{eV}$.  Gravitationally, such dark matter would behave similarly to the axion dark matter.

The stationary initial condition of the scalaron field at the minimum of its potential is a distinctive feature of the described scenario (although, at present, we cannot provide a convincing reason for it, which probably requires going into the poorly controlled Planckian era). If, however, one assumes that the scalaron is excited prior to the electroweak crossover, then its initial energy density may be larger than it is in the described scenario. In this case, the mass $m$ required to ensure the observed amount of dark matter will also be larger. 

Scenario of the type just mentioned was proposed in \cite{Cembranos:2008gj}.  This paper did not take into account the time-dependent shift of the scalaron effective potential caused by its interaction with matter, and simply assumed the scalaron to be ``frozen'' by the Hubble friction at very high temperatures away from the minimum of its proper potential in (\ref{Sgn}). A well-known general problem with scenarios of this type is the necessity of fine-tuning the initial conditions in order to produce the right amount of dark matter \cite{Zlatev:1998yg}. Such a fine-tuning can be relaxed in a special class of models \cite{Zlatev:1998yg, Mishra:2017ehw}, but the model under consideration does not belong to this class. Our scenario avoids the issue of fine tuning by assuming the initial conditions for the scalaron at the minimum of its effective potential. The scalaron is excited by the electroweak crossover, and the abundance of dark matter is controlled only by the value of the fundamental constant $m$.  Also note that our scenario automatically produces adiabatic perturbations in the inhomogeneous universe since the ratio of the energy density of scalaron oscillations to that of radiation is universally the same at the electroweak crossover.

In our scenario, the scalaron always remains in the sub-Planckian region of values $|\phi | \ll M$, so that the issues of non-renormalisability and quantum corrections to the scalaron effective potential are of no crucial importance. For the same reason, the specific form of the $f(R)$ theory in (\ref{Sg}) is also not important; what matters is that the scalaron potential has a minimum with the value of potential corresponding to the current cosmological constant and with mass $m$ of the indicated magnitude.

The quantum scale anomalies arising when proceeding from the Jordan to Einstein frame generate couplings between the scalaron and gauge fields of the form $(\alpha \phi / M)\, \text{tr}\, F_{\mu\nu} F^{\mu\nu}$, where $\alpha = g^2/4 \pi$ is the gauge coupling constant \cite{Cembranos:2008gj, Katsuragawa:2016yir}. They lead to the possibility of scalaron decays into a pair of photons, with lifetime $\tau \sim M^2 / \alpha^2 m^3 \sim 10^{36}\, (\text{eV}/m)^3\, \text{yr}$, far exceeding the age of the universe ($1.4 \times 10^{10}\, \text{yr}$) for our values of the mass $m$.  In this process, parametric resonance in the expanding universe \cite{Kofman:1994rk, Shtanov:1994ce} is not effective because of the small parameter $\alpha^2 \rho_\phi / \rho_\text{total} \ll 1$, which ensures non-adiabatic crossing of the narrow resonance band, thereby giving the usual perturbative expression for the scalaron decay rate (see \cite{Rudenok:2014daa}). One can also show that resonant amplification of the photon production does not take place in scalaron dark-matter halos because of the narrow resonance band and finite halo size. Because of its extremely weak couplings to the usual matter, this light scalaron dark matter appears to be quite ``sterile'' and hard to detect other than gravitationally, unlike potentially detectable dark-matter candidates such as axions or WIMPs.

Since the scalaron was used here to describe dark matter, inflation might be based on the Higgs coupling to the scalar curvature as in \cite{Bezrukov:2007ep}. Such a combined model remains to be elaborated (for inflationary models of this kind, see \cite{Salvio:2015kka, Ema:2017rqn, He:2018gyf, Gorbunov:2018llf, Gundhi:2018wyz, Canko:2019mud, Cheong:2021vdb}).\footnote{Coupling the Higgs field to $f (R)$ gravity can also be used to construct inflationary models with production of primordial black holes constituting dark matter \cite{Pi:2017gih, Gundhi:2020zvb}.}

Finally, it may be worth mentioning that mass scale of the order present in our result (\ref{Om}) also arises in the context of dark energy and neutrino oscillations.  Thus, the dark-energy density can be expressed as
\begin{equation}
\rho_\Lambda = \frac{\Lambda}{8 \pi G} = \left( 2.7 \times 10^{-3}\, \text{eV} \right)^4 \, , 
\end{equation}
while, for the differences of squared masses in solar neutrino oscillations, we have \cite{Zyla:2020zbs}
\begin{equation}
\Delta m^2_\text{sol} \approx \left( 8.7 \times 10^{-3}\, \text{eV} \right)^2 \, .
\end{equation}
At present, this looks like a curious coincidence of mass scales.

\acknowledgments

The author is grateful to Varun Sahni for stimulating discussions.  This work was supported by the National Research Foundation of Ukraine under Project No.~2020.02/0073. 

\appendix

\section{Small scalaron mass}
\label{sec:smass}

Consider the case of the scalaron mass satisfying
\begin{equation} \label{ineq1}
\frac{\lambda v^4}{2 m^2 M^2} \gg 1 \, ,
\end{equation}
which requires $m \ll 5 \times 10^{-6}\, \text{eV}$.  In this case, the electroweak crossover will start from much lower temperature than in the standard cosmology: $T_c \approx {2 m M}/{v}$ [see (\ref{Tc})]. The scalaron in this case resides on the right-wing plateau of its exponential potential; its energy density is $\rho_\phi = V ( \phi_0 ) \approx m^2 M^2 / 2$, while the energy density of radiation is $\rho_\text{rad} \approx 30 T_c^4 \approx 10^3 {m^4 M^4}/{v^4}$. Their ratio is estimated as
\begin{equation}
\frac{\rho_\phi}{\rho_\text{rad}} \simeq 10^{-3} \frac{v^4}{2 m^2 M^2} \, .
\end{equation}
In view of (\ref{ineq1}), the energy density stored in the scalaron is larger or comparable to that of radiation at the time of electroweak crossover. This will lead to an extra period of inflation based on the scalaron, on energy scale much lower than the electroweak energy scale.  Thus, condition (\ref{ineq1}) in the present scenario leads to a very unusual cosmology and, for this reason, is not under consideration here.

\section{Solution for the scalaron}
\label{sec:solution}

Here, we present an approximate solution of equation (\ref{eqxi}) in the case $\xi_h = T^2 / T_c^2$. Given that $T \propto a^{-1}$ and $a \propto t^{1/2}$, equation (\ref{eqxi}) can be written as
\begin{equation}\label{eqxit}
\ddot \xi + \frac{3}{2 t} \dot \xi + m^2 \left( \xi - \frac{t_c}{t} \right) = 0 \, ,
\end{equation}
where $t_c = 1 / 2 H_c$ is the time of electroweak crossover.  Proceeding to the dimensionless variable $x = m t$, we present equation (\ref{eqxit}) in the form
\begin{equation}
\xi'' + \frac{3}{2 x} \xi' + \left( \xi - \frac{\alpha}{x} \right) = 0 \, ,
\end{equation}
where $\alpha = m t_c \gg 1$. The initial condition for this equation is $\xi(\alpha) = 1$, $\xi' (\alpha) = 0$. Introducing the variable 
\begin{equation}
\Delta = \xi - \xi_h = \xi - \frac{\alpha}{x} \, ,
\end{equation}
one then obtains an equation for $\Delta$:
\begin{equation}\label{eqdelta}
\Delta'' + \frac{3}{2 x} \Delta' + \Delta = - \frac{\alpha}{2 x^3} \, ,
\end{equation}
with the initial conditions $\Delta (\alpha) = 0$, $\Delta' (\alpha) = 1/\alpha$.  The right-hand side of this equation initially is of the same order of magnitude as the second term on the left-hand side; however, it decreases much faster with $x$ than the solution of the homogeneous equation and, therefore, to a good approximation, can be neglected. The solution can then be presented in the adiabatic form
\begin{equation}
\Delta (x) = \frac{A}{x^{3/4}} \sin \left( x - \alpha \right) \, ,
\end{equation}
where the initial condition $\Delta (\alpha) = 0$ is imposed.  Another initial condition, $\Delta' (\alpha) = 1/\alpha$, gives the value of the constant: $A = 1 / \alpha^{1/4}$.  As a result, we obtain
\begin{equation}\label{Delta}
\Delta = \frac{1}{\alpha^{1/4} x^{3/4}} \sin \left( x - \alpha \right) 
= \frac{1}{m t_c} \left( \frac{t_c}{t} \right)^{3/4} \sin \left[ m \left( t - t_c \right) \right] 
\end{equation}
and
\begin{equation}\label{xisol}
\xi (t) = \frac{t_c}{t} + \frac{1}{m t_c} \left( \frac{t_c}{t} \right)^{3/4} \sin \left[ m \left( t - t_c \right) \right]  \, .
\end{equation}
As expected, the solution for $\xi$ oscillates with decreasing amplitude around the local minimum $\xi_h = T^2/T_c^2 = t_c/t$. The oscillatory part (\ref{Delta}) has energy (\ref{energy}). This part will formally start dominating in solution (\ref{xisol}) at temperature
\begin{equation} 
T_* \simeq \frac{4 H_c^2}{m^2} T_c \, .
\end{equation}
For the value of $m$ obtained in (\ref{Om}), it is estimated as $T_* \simeq 2 \times 10^{-4}\, T_c \simeq 30\, \text{MeV}$.

\end{document}